\documentclass[article, nonacm]{acmart}
\acmJournal{TSC}

\usepackage{enumitem}
\usepackage{amsmath}
\usepackage{tikz}
\usepackage{pgfplots}
\pgfplotsset{compat=1.18}
\usepackage{pgfplotstable}
\usepackage{booktabs}
\usepackage{tabularx}
\usepgfplotslibrary{statistics, groupplots}
\usepgfplotslibrary{fillbetween}
\usepackage{xcolor}
\definecolor{pptorange}{RGB}{237,125,49}
\definecolor{pptblue}{RGB}{68,114,196}
\definecolor{pptpurple}{RGB}{112,48,160}
\definecolor{pptred}{RGB}{192,0,0}
\definecolor{pptgreen}{RGB}{0, 97, 0} 
\definecolor{pptbrown}{RGB}{153, 102, 51}
\definecolor{pptdarkgrey}{RGB}{64, 64, 64}
\definecolor{pptdarkgreen}{RGB}{0, 112, 33}
\definecolor{pptteal}{RGB}{0,128,192}

\pgfplotsset{every axis/.append style={
    xlabel={$x$},          
    ylabel={$y$},          
    label style={font=\sffamily},
    tick label style={font=\sffamily\scriptsize},
    xticklabel style = {font=\sffamily\scriptsize},
    title style = {font=\normalsize\sffamily},
    ylabel near ticks,
    y label style={font=\sffamily\small},
    xlabel near ticks,
    x label style={font=\sffamily\small},
    },
    }

\begin{document}

\title{Echoes in the Algorithm: Analyzing the Fidelity of User Preferences Against Realized Platform Reach}

\author{Emelia May Hughes}
\email{ehughes8@nd.edu}
\affiliation{%
  \institution{ND-IBM Technology Ethics Lab}
  \institution{University of Notre Dame}
  \state{Indiana}
  \country{USA}
}
\author{Tim Weninger}
\email{tweninger@nd.edu}
\affiliation{%
  \institution{ND-IBM Technology Ethics Lab}
  \institution{University of Notre Dame}
  \state{Indiana}
  \country{USA}
}

\begin{abstract}
What does popular content look like when platforms withhold the usual cues? On TikTok, users still form impressions about which videos are taking off even when likes and view counts are hidden, delayed, or pushed to the margins of the interface. We study this problem through \textit{TokOrNot}, a web-based game in which participants compared pairs of TikTok videos and reported (i) which one they preferred and (ii) which one they believed had reached a larger audience. We benchmark these judgments against verified public view counts, which we use as a bounded proxy for realized platform reach. Across 3,513 judgments from 363 participants, participants identified the higher-reach video only modestly above chance (56.75\%, 95\% CI: 56.01--58.55). Preference aligned with the higher-view video at a similar rate, while preference and prediction matched in 83.48\% of trials (95\% CI: 83.12--85.95). Performance also varied across content categories. Taken together, these results do not suggest that users can reliably read platform success from content alone. Instead, they point to a looser and more uncertain interpretive process in which reach judgments often track personal taste or other weak heuristics when explicit popularity cues are absent. We discuss the implications for algorithmic literacy and for interface designs that reduce visible metrics without leaving users to infer reach from uneven or idiosyncratic cues alone.
\end{abstract}



\maketitle

\section{Introduction}
TikTok is one of the most influential spaces for entertainment, news, and cultural participation today \cite{Matsa-MoreAmericansAre-2023, ZulliZulli-ExtendingInternetMeme-2020, BandyDiakopoulos-TulsaFlopCaseStudy-2020}. Success on the platform matters, but it is not always easy to read. View counts and likes may be hidden, delayed, or simply pushed to the margins of the viewing experience, yet users still form impressions about which videos are taking off and which are not. Creators adjust what they post in response to those impressions. Viewers fold them into how they interpret the For You Page. In both cases, people are not just watching content; they are making sense of what the platform seems to reward.

That sensemaking becomes especially interesting when the usual social signals are reduced. We examine a bounded version of that problem through what we call \emph{outcome legibility}: how users interpret visible signs of platform success when direct indicators of popularity are muted. Our focus is on \emph{realized platform reach}, defined here as the audience a video ultimately attains on-platform and proxied by verified public view counts. The aim is not to recover TikTok’s ranking logic or explain the recommender system from the inside. Instead, we ask a more immediate question from the user’s side of the interface: when popularity cues recede, what do people think successful content looks like?

Work on algorithmic transparency often asks whether users can understand hidden system behavior. Our concern is slightly different. We benchmark human judgments against an observable outcome of platform circulation rather than the recommender’s internal mechanics \cite{EslamiEtAl-AlwaysAssumedThat-2015, DeVitoEtAl-HowPeopleForm-2018}. That distinction matters because users do not wait for transparency to form beliefs about algorithms. They build explanations anyway, drawing from repeated exposure, anecdote, experimentation, and discussion. Prior work describes these beliefs as \emph{folk theories} of algorithms \cite{DeVitoEtAl-AlgorithmsRuinEverything-2017, KarizatEtAl-AlgorithmicFolkTheories-2021}. When explicit counters are absent, those theories may lean less on visible social proof and more on content cues, personal taste, genre expectations, and assumptions about what feels likely to spread.

\begin{figure}[t]
\centering
\includegraphics[width=.4\textwidth]{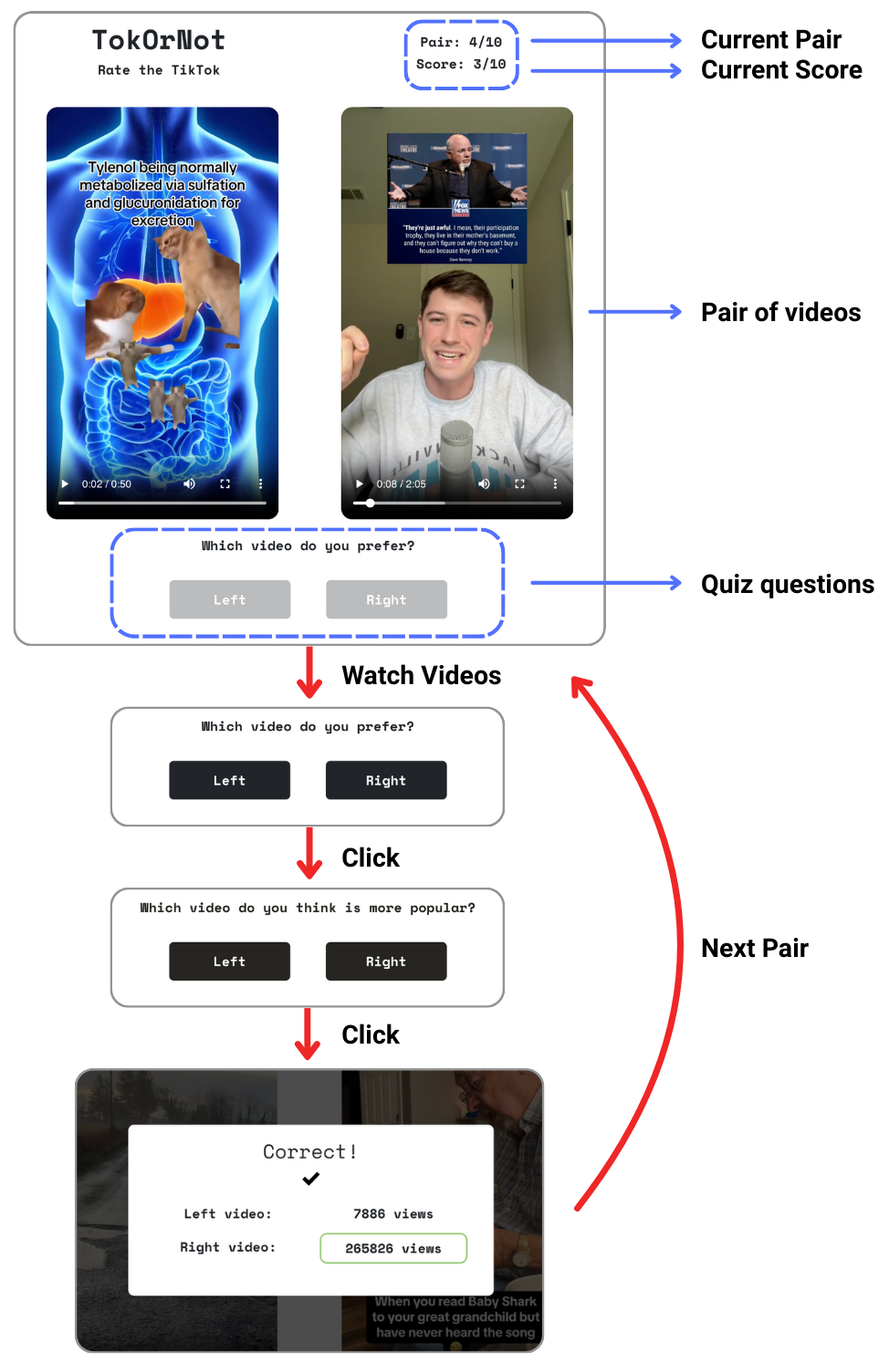}
\caption{TokOrNot Online Game. Participants were shown pairs of TikTok videos and asked to guess which one was more popular based on content alone, without seeing view or like counts. For each pair, they also indicated which video they personally preferred.}
\label{game-flow}
\end{figure}

To study that interpretive process, we built \textit{TokOrNot}, a two-alternative forced-choice game shown in Fig.~\ref{game-flow}. Participants viewed pairs of TikTok videos with all engagement counters hidden and answered two questions for each pair: (1) which video they personally preferred, and (2) which they believed had reached a larger audience. We then compared those judgments against verified public view counts, which we use as a bounded proxy for realized reach \cite{MariaGlenskiEtAl-GuessTheKarmaGameAssess-2018, wit1998rational}. The design adapts ``guess the crowd'' style paradigms to a recommendation-driven short-video environment, where judgments of popularity are likely shaped not only by visible production qualities but also by expectations about trends, genre, and audience fit.

TikTok is a revealing setting for this question because popularity on the platform is never just a matter of raw exposure. Recommendation, imitation, genre convention, and participatory culture all shape what seems likely to travel \cite{KarizatEtAl-AlgorithmicFolkTheories-2021, CollieWilson-Barnao-PlayingTikTokAlgorithmic-2020}. Users learn to read trend formats, production styles, hooks, and affective cues as signs of potential success. At the same time, they also describe the platform as opaque, inconsistent, and difficult to interpret \cite{VelkovaKaun-AlgorithmicResistanceMedia-2021, Darvin-DesignResistancePerformance-2022}. The result is an environment where people routinely make judgments about what spreads widely without having a stable or fully shared basis for doing so.

Those judgments carry consequences beyond momentary curiosity. When a video's apparent success feels surprising, users may read that gap as evidence of suppression, favoritism, or manipulation rather than ordinary variation in circulation. For creators, similar judgments can shape strategy, self-assessment, and emotional investment, especially when people are already trading folk-theory advice about how to gain traction. And for platforms, reducing visible metrics does not remove interpretation; it changes the evidence available for it. Users still try to infer what success looks like, only now from weaker and more uneven cues.

We therefore ask four related questions: (RQ1) how often participants' preferences align with higher-reach content, (RQ2) how accurately they identify the higher-reach video under hidden metrics, (RQ3) whether reach judgments tend to follow preference, and (RQ4) how TikTok's view-based outcome compares with aggregated human judgments. Across 3,513 trials from 363 participants, participants identified the higher-reach video only modestly above chance (56.75\%, 95\% CI: 56.01--58.55). Preference aligned with the higher-view video at a similar rate, while preference and prediction matched in 83.48\% of trials (95\% CI: 83.12--85.95). Performance also varied across content categories. Taken together, these results do not suggest that users can reliably read platform success from content alone. Instead, they point to a looser and more uncertain interpretive process in which judgments about reach often track personal taste or other weak heuristics when explicit popularity cues are absent.

\textbf{Contributions.}
This paper makes two main contributions. First, it provides an empirical benchmark of how people infer relative reach in a metric-hidden, pairwise TikTok task using \textit{TokOrNot}, with materials released to support replication. Second, it uses an outcome-legibility framing to study how users interpret platform success when obvious popularity cues are reduced, and to motivate interface support that helps users orient to reach outcomes without simply restoring precise counters.

\section{Related Work}

\paragraph{Algorithmic sensemaking and folk theories.}
People do not need full visibility into a platform's ranking system to form beliefs about how it works. Across social media, users develop informal explanations for why certain content appears, spreads, or disappears, then use those explanations to guide their own behavior \cite{DeVitoEtAl-AlgorithmsRuinEverything-2017, DeVitoEtAl-HowPeopleForm-2018, EslamiEtAl-AlwaysAssumedThat-2015}. Work on \emph{algorithmic folk theories} shows that these beliefs are not merely private guesses. They are revised through repeated use, social discussion, and observation of platform outcomes \cite{KarizatEtAl-AlgorithmicFolkTheories-2021, MarcuEtAl-DesigningCollaborativeReflection-2014, MuralikumarBietz-VisualizingAlgorithmicSelection-2019, VaccaroEtAl-ContestabilityAlgorithmicSystems-2019, MaywormEtAl-ContentModerationFolkTheories-2024, DelmonacoEtAl-WhatAreYouDoingTikTok-2024}. Recent work on marginalized users makes this especially visible by showing how folk theories develop around content moderation and shadowbanning, including collaborative efforts to test or ``prove'' suppression \cite{MaywormEtAl-ContentModerationFolkTheories-2024, DelmonacoEtAl-WhatAreYouDoingTikTok-2024}. That perspective is important here because our question is not whether participants can recover TikTok's hidden ranking logic. It is whether, in the absence of explicit counters, they can read a visible \emph{outcome} of platform circulation: which video ultimately reached the larger audience.

\paragraph{Popularity cues, social influence, and reduced metrics.}
A long line of research shows that visible popularity signals do not simply reflect collective judgment; they can actively shape it. Early advantages can snowball into herding, inequality, and unstable success, making popularity partly a product of social influence rather than an independent measure of quality \cite{SalganikEtAl-ExperimentalStudyInequality-2006, LorenzEtAl-HowSocialInfluence-2011, WangEtAl-QuantifyingHerdingEffects-2014, LevMuchnikEtAl-SocialInfluenceBias-2013, WuHuberman-HowPublicOpinion-2008, JacoviEtAl-PerceptionOthersInferring-2014}. One response has been to hide, delay, or otherwise soften visible metrics. But reducing counters does not eliminate interpretation; it changes the evidence available for it. Recent visibility scholarship similarly suggests that users reason not only about what is seen by others, but also about how they themselves are rendered visible to platforms and algorithms \cite{BartaAndalibi-TheorizingSelfVisibility-2024}. When users can no longer lean on likes or views directly, they may infer popularity from weaker signals such as polish, genre familiarity, affect, or resemblance to already-legible trend formats. Our study speaks to that shift in the interpretive environment rather than to the direct causal effects of metric-hiding policies themselves.

\paragraph{TikTok, recommendation, and participatory cues of success.}
TikTok makes this problem especially sharp because judgments of popularity are entangled with recommendation and participatory culture. Prior work shows that users build folk theories about the For You Page, develop expectations about what kinds of content the platform rewards, and read platform behavior through repeated encounters with trends, repetition, and personalization \cite{CollieWilson-Barnao-PlayingTikTokAlgorithmic-2020, KarizatEtAl-AlgorithmicFolkTheories-2021}. More recent work sharpens that picture from several angles. Audit-oriented work shows how TikTok recommendations balance exploration and exploitation and identifies behavioral factors such as watch time, likes, and following as drivers of personalization \cite{VombatkereEtAl-TikTokArtPersonalization-2024}. Interview studies of the For You Page likewise show that users develop tactics and folk theories to explain and steer unwanted recommendations \cite{VeraGhosh-ControllingUnwantedContentTikTok-2025}. Creator-centered research further shows how algorithmic expectations shape strategy, emotional labor, and adaptation on creator platforms \cite{ChoiEtAl-CreatorFriendlyAlgorithms-2023}, while TikTok-specific work on algospeak highlights how users alter expression in response to perceived moderation and ranking dynamics \cite{KlugEtAl-AlgorithmAwarenessAlgospeak-2023}. In a metric-hidden setting, users may therefore treat cultural fit as a cue to reach just as readily as they treat explicit signs of popularity. That possibility matters for interpreting errors in our task: miscalibration may reflect not random guessing, but systematic reliance on genre-shaped expectations about what should do well on TikTok.

\paragraph{Guess-the-crowd paradigms and our contribution.}
Methodologically, our study is closest to ``guess the crowd'' and social-rating paradigms that ask participants to predict collective outcomes from limited information. GuessTheKarma \cite{MariaGlenskiEtAl-GuessTheKarmaGameAssess-2018}, for example, used pairwise comparisons to examine whether users could anticipate Reddit popularity better than chance, while Condorcet-style approaches similarly treat aggregate judgment as something that can be elicited through comparative choice \cite{wit1998rational}. \textit{TokOrNot} builds on that tradition rather than claiming a wholly new experimental paradigm. The difference lies in the setting and the target of inference. We move the task into a recommendation-driven short-video environment where popularity is shaped by platform-specific conventions, and we benchmark judgments against realized platform reach proxied by public view counts rather than against a generic crowd preference signal. This lets us ask a narrower question than the broader outcome-legibility frame might suggest: when obvious metrics are unavailable, how well can people infer \emph{relative reach}, and how often do those judgments collapse into personal preference instead?

\section{Methodology}

This study used a preregistered online behavioral experiment built around repeated pairwise judgments. Participants completed \textit{TokOrNot}, a web-based task in which they compared pairs of TikTok videos with all engagement metrics hidden. For each pair, they indicated which video they personally preferred and which they believed had reached a larger audience on TikTok. All procedures were approved by the [\textit{redacted}] Institutional Review Board and followed standard ethical practices for online behavioral research. Preregistration details, data, and evaluation scripts are available at \url{https://osf.io/gjzpc/overview?view_only=5b431e7a85a24ae2a36564d12117a0cd}.

We treat \textit{TokOrNot} as a study instrument for eliciting comparative judgments under hidden metrics rather than as a simulation of the TikTok For You Page. TikTok is ordinarily encountered as a sequential feed, not a side-by-side comparison task. The paired design therefore serves a narrower methodological purpose: it holds exposure constant across two candidate videos and makes participants' judgments legible in a controlled, guess-the-crowd style format. This lets us examine how people infer \emph{relative reach} when explicit popularity cues are unavailable.

\subsection{Video and Pair Collection}

We curated TikTok videos spanning a broad range of public view counts and content types. Metadata was retrieved through the TikTok Research API and then manually verified and corrected where needed. Videos were grouped into five wide view-count buckets (\emph{very low}, \emph{low}, \emph{medium}, \emph{high}, and \emph{viral}) to reduce sensitivity to small fluctuations and support interpretable comparisons between videos with meaningfully different observed reach outcomes.

Rather than pairing videos across unrelated genres, we constructed comparisons within category and across view-count buckets. This does not make the task naturalistic in the sense of reproducing the For You Page. It does, however, make the comparisons more interpretable by reducing a simple cross-genre confound in which participants might choose between, for example, an animal video and a dance video on the basis of broad taste alone. The analysis uses nine content categories: Animals, Comedy, Dancing, Education, Meme, Random, Singing, Sketch Comedy, and Trend.

\subsection{Measures}

\textbf{Realized platform reach.}
We use verified public view counts as a bounded proxy for realized on-platform reach and treat the higher-view item within a pair as the pair-level reference for ``higher reach.'' This proxy captures a publicly recorded attention outcome on the platform, but it is not a direct measure of algorithmic exposure alone. View totals may also reflect creator audience, video age, and off-platform traffic. The paper therefore benchmarks human judgments against an observable reach outcome, not against TikTok's internal ranking logic.

\textbf{Prediction accuracy.}
For each trial, \emph{accuracy} equals 1 when a participant correctly identified the higher-view video in the pair and 0 otherwise.

\textbf{Preference--reach alignment.}
For each trial, \emph{preference--reach alignment} equals 1 when the participant preferred the higher-view video and 0 otherwise. This is the quantity reported in the analysis as \texttt{pref\_popular}.

\textbf{Preference substitution.}
For each trial, \emph{substitution} equals 1 when the participant's preference choice matched their reach choice and 0 otherwise. We use this term descriptively to indicate alignment between the two judgments within the task. Because preference was always elicited before reach prediction, this should not be treated as clean evidence of a naturally occurring heuristic independent of task order.

\paragraph{Assumptions and stabilization.}
Interpreting public view counts as realized platform reach rests on three bounded assumptions. First, \emph{monotonicity}: conditional on content and timing, greater exposure should generally increase view totals. Second, \emph{stabilization}: by the time metadata was sampled and verified, view counts had largely stabilized relative to the comparisons being made. Third, \emph{comparable opportunity within pairs}: pairing within category and across wide reach buckets creates a more interpretable contrast than comparing arbitrary videos with little shared context. These assumptions are intended to support within-pair comparisons of an observable attention outcome; they do not justify causal claims about treatment by the recommender.

\begin{figure*}[t]
\centering
\includegraphics[width=.95\textwidth]{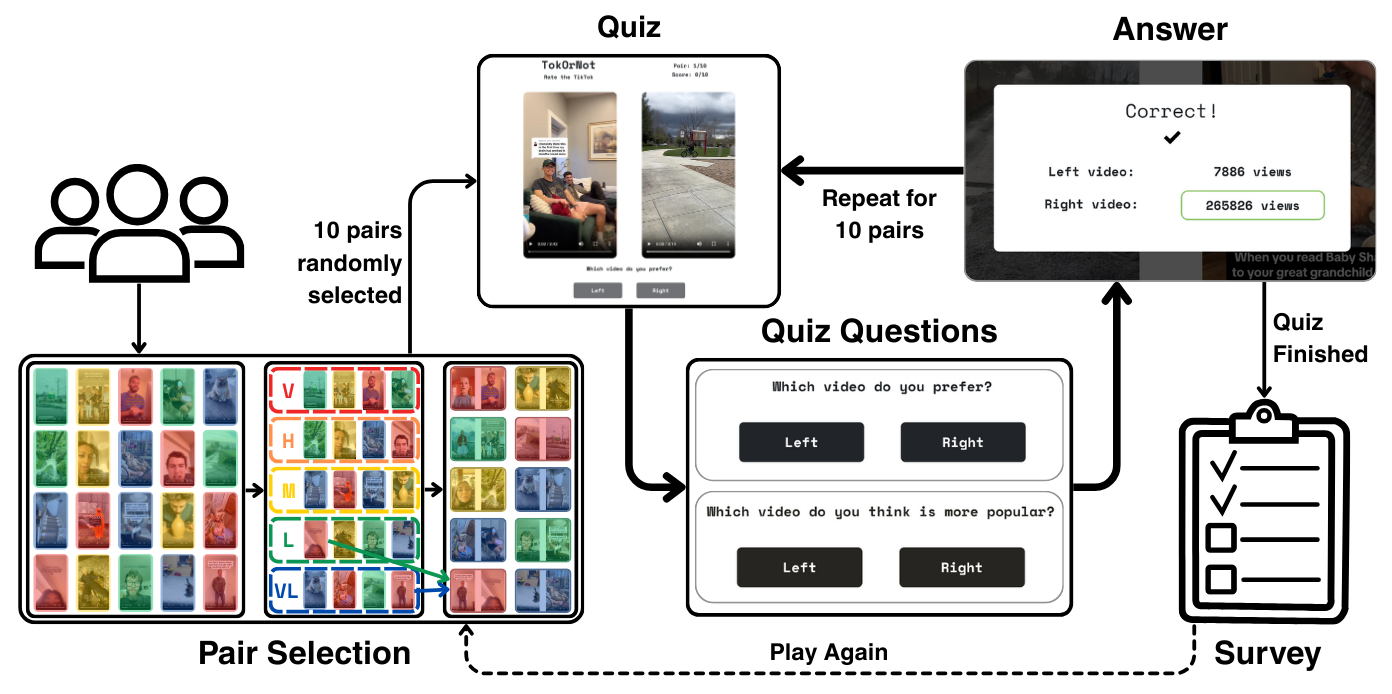}
\caption{TokOrNot experimental methodology. Pairs were drawn from five view-count buckets and displayed side-by-side to participants. After watching both videos, participants answered two questions: which video they personally preferred, and which they believed had reached a larger audience on TikTok. Each participant completed up to ten trials with all engagement metrics hidden, followed by a short survey about TikTok usage and perceived popularity cues.}
\label{fig:method}
\end{figure*}

\subsection{Participants and Procedure}

Participants were recruited on Prolific and completed \textit{TokOrNot} online. The final analysis dataset includes $N = 363$ participants contributing $N = 3513$ trial-level judgments across $N = 1978$ unique video pairs. Participants were U.S.-based adults. Ages ranged from 18 to 82 years ($M = 43.6$), with a near-even gender split.

Data collection occurred in two Prolific batches (310 and 216 recruits), each preceded by a 10-participant pilot. In the second batch, pair selection was briefly adjusted to prioritize pairs with fewer than four existing responses in order to improve coverage across under-rated pairs and categories, after which assignment returned to random sampling.

Each participant completed up to ten trials. In each trial, two within-category TikTok videos were presented side by side with all engagement counters hidden. Videos could be replayed as needed. To ensure exposure before response, participants were required to watch at least 90\% of each video; for videos longer than 40 seconds, 30 seconds of watch time was sufficient before the response buttons were enabled.

After viewing both videos, participants answered two forced-choice questions: which video they personally preferred, and which video they believed had reached a larger audience on TikTok. After each trial, the interface displayed the verified view counts for both videos and indicated whether the reach judgment was correct. A progress tracker also showed task completion and current accuracy. After the ten trials, participants completed a short survey about TikTok usage and perceived popularity cues.

One design choice is especially important for interpretation. The within-trial question order was fixed: participants always reported preference before reach prediction. We retain that design as part of the instrument as deployed, but we treat it as a limitation rather than as a neutral feature of the task. The analyses therefore use trial-order terms and early-trial robustness checks to bound possible across-trial learning or feedback drift, while remaining conservative about any claim that the substitution result reflects a stable heuristic rather than within-trial anchoring or consistency pressure.

\subsection{Data Analysis}

Our analyses operate at the trial level, where each observation consists of one participant's preference and reach judgment for one video pair. Descriptive statistics provide the primary answers to RQ1--RQ3: how often preference aligns with higher reach, how accurately participants identify the higher-reach video, and how often preference matches prediction. The regression models play a narrower role. They test whether the main patterns are robust to structural factors such as reach gap, trial order, and content category, rather than serving as the sole basis for the paper's claims.

\paragraph{Primary inference model.}
We fit a logistic regression predicting whether the participant correctly identified the higher-reach video:
{\small
\begin{equation}
\begin{aligned}
\text{logit}\big(\Pr(\text{correct}=1)\big) = \beta_0 
&+ \beta_1 \cdot \text{abs\_reach\_gap\_c} \\
&+ \beta_2 \cdot \text{trial\_index\_c}
+ \sum_j \beta_j \cdot \text{category}_j .
\end{aligned}
\end{equation}
}

Here, \texttt{abs\_reach\_gap} is the absolute difference between the two videos' log-transformed view counts, and \texttt{abs\_reach\_gap\_c} is its mean-centered version. \texttt{trial\_index\_c} is the mean-centered trial order.

\paragraph{Substitution model.}
To examine whether participants' reach judgments tracked their own preferences, we fit a parallel logistic regression predicting substitution:
{\small
\begin{equation}
\begin{aligned}
\text{logit}\big(\Pr(\text{substitution}=1)\big) = \beta_0 
&+ \beta_1 \cdot \text{abs\_reach\_gap\_c} \\
&+ \beta_2 \cdot \text{trial\_index\_c}
+ \sum_j \beta_j \cdot \text{category}_j .
\end{aligned}
\end{equation}
}

\paragraph{Preference-conditioning comparison.}
To assess whether preference--reach alignment related to prediction accuracy, we compared two models: one including \texttt{pref\_popular} as a predictor and one without. Here, \texttt{pref\_popular} indicates whether the participant preferred the higher-view video in the pair. Model comparison was based on Akaike Information Criterion (AIC).

\paragraph{Inference and robustness.}
All inferential models were fit as logistic generalized linear models with two-way clustered standard errors, clustering on both \texttt{participant\_id} and \texttt{pair\_id}. This accounts for repeated judgments by the same participant and repeated ratings of the same video pair. The primary model included 363 participant clusters and 1978 pair clusters. Headline rates are reported with participant-cluster bootstrap confidence intervals (2000 resamples). We also report robustness checks using alternative specifications, including signed reach gap, exclusion of the first trial, and early-trial subsets.

\paragraph{RQ4 analysis.}
For RQ4, we use Survey Equivalence to compare TikTok's view-based signal to aggregated human judgments. In this framework, the view-based outcome is treated as an additional classifier and compared against groups of human raters of varying size. This lets us translate the agreement level into a more interpretable unit: how many independent human raters provide a signal comparable to the view-based benchmark.

\section{Results}

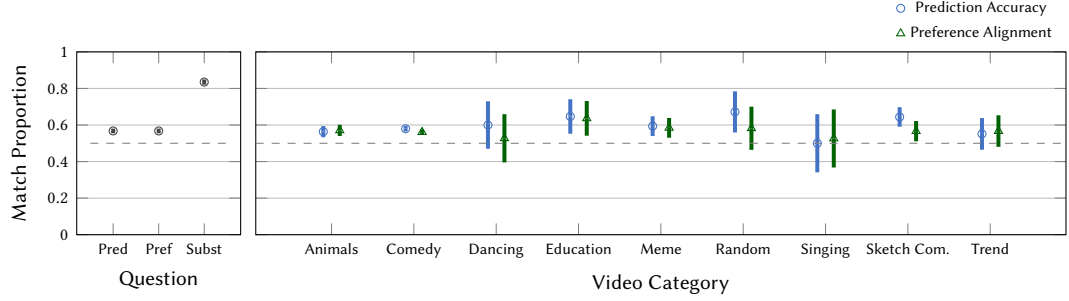
\begin{figure*}[t]
    \centering
\pgfplotstableread[col sep=comma]{
x,Metric,Rate,CI
0,Prediction,0.5675,0.0127
1,Preference,0.5675,0.0127
2,Substitution,0.8348,0.0142
}\matchrates

\pgfplotstableread[col sep=comma]{
x,Category,PredictionAccuracy,PredictionAccuracyCI,PreferenceAlignment,PreferenceAlignmentCI,PredictionPreferenceMatch,PredictionPreferenceMatchCI
0,Animals,0.5635,0.0305,0.5704,0.0305,0.8670,0.0209
1,Comedy,0.5797,0.0153,0.5609,0.0154,0.8462,0.0112
2,Dancing,0.6000,0.1295,0.5273,0.1319,0.7455,0.1151
3,Education,0.6465,0.0942,0.6364,0.0948,0.8283,0.0743
4,Meme,0.5938,0.0538,0.5844,0.0540,0.8406,0.0401
5,Random,0.6716,0.1125,0.5821,0.1181,0.9104,0.0684
6,Singing,0.5000,0.1590,0.5263,0.1588,0.8158,0.1233
7,Sketch Comedy,0.6440,0.0534,0.5663,0.0553,0.8252,0.0423
8,Trend,0.5512,0.0865,0.5669,0.0862,0.8268,0.0658
}\categorydata

\begin{tikzpicture}
\begin{groupplot}[
    group style={
        group size=2 by 1,
        horizontal sep=0.2cm,
        vertical sep=0.2cm,
    }, 
    ymajorgrids=true,
]

\nextgroupplot[
    enlarge x limits=0.4,
    ylabel={Match Proportion},
    ymin=0, ymax=1,
    height=4cm,
    width=3.75cm,
    xtick={0,1,2},
    xticklabels={Pred, Pref, Subst},
    xlabel={Question},
]

\addplot+[
    only marks,
    mark=o,
    mark options={fill=pptdarkgrey, draw=pptdarkgrey},
    mark size=1.5pt,
    error bars/.cd,
        y dir=both,
        y explicit,
        error mark=bar,
        error bar style={line width=1.5pt, draw=pptdarkgrey},
] table [
    x expr=\coordindex, 
    y=Rate, 
    y error=CI,
] {\matchrates};

\draw[dashed, gray] ({axis cs:-0.5,0.5}) -- ({axis cs:2.5,0.5});

\nextgroupplot[
    width=12.3cm,
    height=4cm,
    ymin=0.0, ymax=1.0,
    enlarge x limits=0.1,
    xticklabels={Animals,Comedy,Dancing,Education,Meme,Random,Singing,Sketch Com.,Trend},
    xtick={0,1,2,3,4,5,6,7,8},
    ylabel={},
    ytick={},
    yticklabels={},
    xlabel={Video Category},
    legend style={at={(1.0,1.0)}, anchor=south east, legend columns=1, draw=none, fill=none, font=\sffamily\scriptsize},
]

\addplot+[
    only marks,
    mark=o,
    mark options={fill=pptblue, draw=pptblue},
    mark size=1.5pt,
    error bars/.cd,
        y dir=both,
        y explicit,
        error mark=bar,
        error bar style={line width=1.5pt, draw=pptblue},
] table [
    x expr=\coordindex - 0.1, 
    y=PredictionAccuracy, 
    y error=PredictionAccuracyCI,
] {\categorydata};

\addplot+[
    only marks,
    mark=triangle,
    mark options={fill=pptgreen, draw=pptgreen},
    mark size=1.75pt,
    error bars/.cd,
        y dir=both,
        y explicit,
        error mark=bar,
        error bar style={line width=1.5pt, draw=pptgreen},
]
table[x expr=\coordindex + 0.1, y=PreferenceAlignment, y error=PreferenceAlignmentCI] {\categorydata};


\legend{Prediction Accuracy, Preference Alignment}

\draw[dashed,gray] ({rel axis cs:0,0.5}) -- ({rel axis cs:1,0.5});

\end{groupplot}
\end{tikzpicture}
    \caption{Participant performance across TikTok video comparisons. \textbf{Left:} Overall match rates showing prediction accuracy (correct identification of the higher-reach video), preference--reach alignment (preferring the higher-reach video), and preference--prediction matching. Error bars indicate 95\% participant-cluster bootstrap confidence intervals; the dashed line denotes chance level (50\%). \textbf{Right:} Match rates broken down by content category, revealing variation in performance and judgment alignment.}
    \label{fig:overall_stats}
\end{figure*}

\subsection{Accuracy, Preference--Reach Alignment, and Preference--Prediction Matching (RQ1--RQ3)}

We begin with the headline rates for recognition, preference--reach alignment, and preference--prediction matching. Across $N=3513$ trials from $N=363$ participants, participants identified the higher-reach video only modestly above chance: 56.75\% (95\% CI: 56.01--58.55; participant-cluster bootstrap). Preference aligned with the higher-view video at the same aggregate rate, 56.75\% (95\% CI: 56.01--58.55), while preference and reach prediction matched in 83.48\% of trials (95\% CI: 83.12--85.95) (Fig.~\ref{fig:overall_stats}).

The strongest pattern here is not the equality of the first two percentages, but the much higher rate of preference--prediction matching. Because the task elicited both judgments on the same pair, that 83.48\% figure is best interpreted as strong within-task alignment between preference and reach prediction. It is consistent with participants often carrying their own tastes into reach judgments under hidden metrics, but, given the fixed question order, it should not be treated as definitive evidence of a naturally occurring heuristic independent of the task structure.

Performance varied across content categories more in pattern than in overall inferential strength (Fig.~\ref{fig:overall_stats}, right). Accuracy ranged from 50.00\% for \textit{Singing} to 67.16\% for \textit{Random}. Preference--prediction matching was highest in \textit{Random} (91.04\%) and lowest in \textit{Dancing} (74.55\%). At the same time, the overall association between category and correctness was not statistically significant ($\chi^2 = 12.10$, df = 8, $p = .15$), which is consistent with the category terms in the primary model being small and non-significant (Table~\ref{tab:primary_model}). We therefore treat these category differences as descriptive structure rather than strong evidence of systematic category effects.

\begin{figure}[t]
    \centering
    \pgfplotstableread[col sep=comma]{
firstlevel,secondlevel,accuracy,n,se,ci95
0,1,0.375,8,0.1711632992203644,0.33548006647191425
0,2,0.5833333333333334,12,0.14231876063832777,0.2789447708511224
0,3,0.5116279069767442,43,0.0762286633173339,0.14940818010197443
0,4,0.5074626865671642,67,0.06107791797627373,0.11971271923349651
}\categorydataverylow

\pgfplotstableread[col sep=comma]{
firstlevel,secondlevel,accuracy,n,se,ci95
1,0,0.375,8,0.1711632992203644,0.33548006647191425
1,1,0.25,4,0.21650635094610965,0.4243524478543749
1,2,0.5238095238095238,42,0.07706415178216067,0.1510457374930349
1,3,0.6842105263157895,95,0.047690502419185245,0.09347338474160308
1,4,0.59375,288,0.028940248399600087,0.05672288686321617
}\categorydatalow

\pgfplotstableread[col sep=comma]{
firstlevel,secondlevel,accuracy,n,se,ci95
2,0,0.5833333333333334,12,0.14231876063832777,0.2789447708511224
2,1,0.5238095238095238,42,0.07706415178216067,0.1510457374930349
2,2,0.4810126582278481,79,0.05621381883942542,0.11017908492527381
2,3,0.6046511627906976,301,0.028181196702704588,0.05523514553730099
2,4,0.616519174041298,1017,0.015247008574882178,0.029884136806769067
}\categorydatamedium

\pgfplotstableread[col sep=comma]{
firstlevel,secondlevel,accuracy,n,se,ci95
3,0,0.5116279069767442,43,0.0762286633173339,0.14940818010197443
3,1,0.6842105263157895,95,0.047690502419185245,0.09347338474160308
3,2,0.6046511627906976,301,0.028181196702704588,0.05523514553730099
3,3,0.5454545454545454,286,0.029443194346877078,0.05770866091987907
3,4,0.5737327188940092,1736,0.011869186332522496,0.02326360521174409
}\categorydatahigh

\pgfplotstableread[col sep=comma]{
firstlevel,secondlevel,accuracy,n,se,ci95
4,0,0.5074626865671642,67,0.06107791797627373,0.11971271923349651
4,1,0.59375,288,0.028940248399600087,0.05672288686321617
4,2,0.616519174041298,1017,0.015247008574882178,0.029884136806769067
4,3,0.5737327188940092,1736,0.011869186332522496,0.02326360521174409
4,4,0.5783014236622485,2037,0.010941646602065407,0.021445627340048196
}\categorydataviral

\pgfplotstableread[col sep=comma]{
x,secondlevel,accuracy,n,se,ci95
0,1,0.5,8,0.1767766952966369,0.3464823227814083
0,2,0.6666666666666666,12,0.13608276348795434,0.2667222164363905
0,3,0.46511627906976744,43,0.07606348725472868,0.1490844350192682
0,4,0.5522388059701493,67,0.06075042007726675,0.11907082335144283
}\categorydataprefverylow

\pgfplotstableread[col sep=comma]{
x,secondlevel,accuracy,n,se,ci95
1,0,0.5,8,0.1767766952966369,0.3464823227814083
1,1,0.5,4,0.25,0.49
1,2,0.4523809523809524,42,0.07680098373375653,0.1505299281181628
1,3,0.5894736842105263,95,0.05047088372035536,0.09892293209189651
1,4,0.5520833333333334,288,0.029302500745865014,0.057432901461895426
}\categorydatapreflow

\pgfplotstableread[col sep=comma]{
x,secondlevel,accuracy,n,se,ci95
2,0,0.6666666666666666,12,0.13608276348795434,0.2667222164363905
2,1,0.4523809523809524,42,0.07680098373375653,0.1505299281181628
2,2,0.45569620253164556,79,0.05603312464373454,0.1098249243017197
2,3,0.5946843853820598,301,0.02829806158704382,0.055464200710605885
2,4,0.5870206489675516,1017,0.01543939866459039,0.030261221382597164
}\categorydataprefmedium

\pgfplotstableread[col sep=comma]{
x,secondlevel,accuracy,n,se,ci95
3,0,0.46511627906976744,43,0.07606348725472868,0.1490844350192682
3,1,0.5894736842105263,95,0.05047088372035536,0.09892293209189651
3,2,0.5946843853820598,301,0.02829806158704382,0.055464200710605885
3,3,0.5209790209790209,286,0.029539583534306742,0.057897583727241214
3,4,0.5662442396313364,1736,0.011894595176863592,0.02331340654665264
}\categorydataprefhigh

\pgfplotstableread[col sep=comma]{
x,secondlevel,accuracy,n,se,ci95
4,0,0.5522388059701493,67,0.06075042007726675,0.11907082335144283
4,1,0.5520833333333334,288,0.029302500745865014,0.057432901461895426
4,2,0.5870206489675516,1017,0.01543939866459039,0.030261221382597164
4,3,0.5662442396313364,1736,0.011894595176863592,0.02331340654665264
4,4,0.5645557191948944,2037,0.010985610240716608,0.02153179607180455
}\categorydataprefviral

\begin{tikzpicture}
\begin{groupplot}[
    group style={
        group size=1 by 2,
        horizontal sep=0.2cm,
        vertical sep=0.2cm,
    }, 
    width=8.5cm,
    height=3.75cm,
    ymajorgrids=true,
    ymin=0.0, ymax=1.0,
    enlarge x limits=0.15,
    ytick={0,.5,1.0},
    yticklabels={0,0.5,1.0},
    xticklabel style={font=\small},
    yticklabel style={font=\small},
    xlabel style={font=\small},
    ylabel style={align=center, font=\small\sffamily},
]

\nextgroupplot[
    xticklabels={},
    xtick={0,1,2,3,4},
    yticklabel style={font=\small},
    ylabel style={align=center},
    ylabel={Prediction \\ Match \%},
    xlabel={},
    legend style={at={(1.0,1.05)}, anchor=south east, legend columns=6, draw=none, fill=none, font=\sffamily\scriptsize},
]

\addlegendimage{empty legend}

\addplot+[
    only marks,
    mark=o,
    mark options={fill=pptred, draw=pptred},
    mark size=1.75pt,
    error bars/.cd,
        y dir=both,
        y explicit,
        error mark=bar,
        error bar style={line width=1.5pt, draw=pptred},
]
table[x expr=\coordindex - 0.2, y=accuracy, y error=ci95] {\categorydataverylow};

\addplot+[
    only marks,
    mark=triangle,
    mark options={fill=pptgreen, draw=pptgreen},
    mark size=1.75pt,
    error bars/.cd,
        y dir=both,
        y explicit,
        error mark=bar,
        error bar style={line width=1.5pt, draw=pptgreen},
]
table[x expr=\coordindex - 0.1, y=accuracy, y error=ci95] {\categorydatalow};

\addplot+[
    only marks,
    mark=square,
    mark options={fill=pptteal, draw=pptteal},
    mark size=1.75pt,
    error bars/.cd,
        y dir=both,
        y explicit,
        error mark=bar,
        error bar style={line width=1.5pt, draw=pptteal},
]
table[x expr=\coordindex - 0.0, y=accuracy, y error=ci95] {\categorydatamedium};

\addplot+[
    only marks,
    mark=diamond,
    mark options={fill=pptpurple, draw=pptpurple},
    mark size=1.75pt,
    error bars/.cd,
        y dir=both,
        y explicit,
        error mark=bar,
        error bar style={line width=1.5pt, draw=pptpurple},
]
table[x expr=\coordindex + 0.1, y=accuracy, y error=ci95] {\categorydatahigh};

\addplot+[
    only marks,
    mark=triangle*,
    mark options={fill=pptbrown, draw=pptbrown},
    mark size=1.75pt,
    error bars/.cd,
        y dir=both,
        y explicit,
        error mark=bar,
        error bar style={line width=1.5pt, draw=pptbrown},
]
table[x expr=\coordindex + 0.2, y=accuracy, y error=ci95] {\categorydataviral};


\addlegendentry{\textbf{Right Video Views: }}
\addlegendentry{very low}
\addlegendentry{low}
\addlegendentry{medium}
\addlegendentry{high}
\addlegendentry{viral}

\draw[dashed,gray] ({rel axis cs:0,0.5}) -- ({rel axis cs:1,0.5});

\nextgroupplot[
    xticklabels={very low, low, medium, high, viral},
    xtick={0,1,2,3,4},
    xticklabel style={font=\small},
    yticklabel style={font=\small},
    xlabel style={font=\small},
    ylabel style={align=center, font=\small\sffamily},
    ylabel={Preference \\ Match \%},  
    xlabel={Left Video Views},
]

\addplot+[
    only marks,
    mark=o,
    mark options={fill=pptred, draw=pptred},
    mark size=1.75pt,
    error bars/.cd,
        y dir=both,
        y explicit,
        error mark=bar,
        error bar style={line width=1.5pt, draw=pptred},
]
table[x expr=\coordindex - 0.2, y=accuracy, y error=ci95] {\categorydataprefverylow};

\addplot+[
    only marks,
    mark=triangle,
    mark options={fill=pptgreen, draw=pptgreen},
    mark size=1.75pt,
    error bars/.cd,
        y dir=both,
        y explicit,
        error mark=bar,
        error bar style={line width=1.5pt, draw=pptgreen},
]
table[x expr=\coordindex - 0.1, y=accuracy, y error=ci95] {\categorydatapreflow};

\addplot+[
    only marks,
    mark=square,
    mark options={fill=pptteal, draw=pptteal},
    mark size=1.75pt,
    error bars/.cd,
        y dir=both,
        y explicit,
        error mark=bar,
        error bar style={line width=1.5pt, draw=pptteal},
]
table[x expr=\coordindex - 0.0, y=accuracy, y error=ci95] {\categorydataprefmedium};

\addplot+[
    only marks,
    mark=diamond,
    mark options={fill=pptpurple, draw=pptpurple},
    mark size=1.75pt,
    error bars/.cd,
        y dir=both,
        y explicit,
        error mark=bar,
        error bar style={line width=1.5pt, draw=pptpurple},
]
table[x expr=\coordindex + 0.1, y=accuracy, y error=ci95] {\categorydataprefhigh};

\addplot+[
    only marks,
    mark=triangle*,
    mark options={fill=pptbrown, draw=pptbrown},
    mark size=1.75pt,
    error bars/.cd,
        y dir=both,
        y explicit,
        error mark=bar,
        error bar style={line width=1.5pt, draw=pptbrown},
]
table[x expr=\coordindex + 0.2, y=accuracy, y error=ci95] {\categorydataprefviral};


\draw[dashed,gray] ({rel axis cs:0,0.5}) -- ({rel axis cs:1,0.5});

\end{groupplot}
\end{tikzpicture}
    \caption{Prediction and preference--reach alignment by view-count bucket (popularity tier). \textbf{Top:} Proportion of trials where participants correctly identified the higher-reach video. \textbf{Bottom:} Proportion of trials where participants preferred the higher-reach video. Error bars show 95\% confidence intervals.}
    \label{fig:pair_accuracy}
\end{figure}
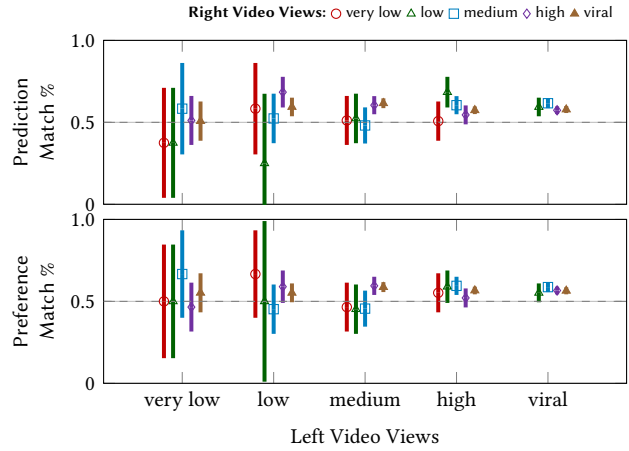

Across bucket contrasts (Fig.~\ref{fig:pair_accuracy}), participants were descriptively more accurate when pairs were separated by larger reach gaps. For example, accuracy was 68.4\% for \textit{high--low} pairs and 61.7\% for \textit{medium--viral} pairs. However, the inferential evidence for a reach-gap effect remained weak (chi-squared $p=.063$; bucket-contrast regression $\beta=0.044$, $p=.111$), and the primary model's absolute gap term was non-significant ($\beta=-0.006$, $p=.609$). Taken together, these descriptive patterns suggest that larger contrasts may sometimes be easier to judge, but they do not provide strong evidence that gap size reliably improves accuracy in this dataset.

\subsection{Modeling Correctness and Preference--Prediction Matching (RQ2--RQ3)}

We next modeled correctness and preference--prediction matching using logistic regression with two-way clustered standard errors by participant and pair (Methods). In the primary model predicting correctness, the absolute reach gap term was small and non-significant ($\beta=-0.006$, $p=.609$), and trial index was also non-significant ($p=.123$), indicating limited evidence that recognition improved over the course of the task (Table~\ref{tab:primary_model}).

In the parallel model predicting preference--prediction matching, reach gap and trial index were likewise non-significant ($p=.357$ and $p=.992$). The absence of a trial-index effect matters because it suggests that the high preference--prediction match rate is not well explained by gradual learning or accumulating feedback across trials. At the same time, these models do not eliminate the possibility of \emph{within-trial} anchoring, since preference was always elicited immediately before the reach judgment.

To assess how strongly preference--reach alignment related to prediction accuracy, we compared a baseline correctness model to a version that also included \texttt{pref\_popular}. Including \texttt{pref\_popular} substantially improved fit ($\Delta$AIC $= -1800.47$), indicating that whether a participant preferred the higher-view video was strongly associated with whether they correctly identified the higher-reach item. This result should be read as evidence that preference alignment carries substantial explanatory power for correctness in this task, not as proof that preference independently determines reach inference outside the experimental setting.

Robustness checks (Table~\ref{tab:robustness}) yielded the same overall picture. The absolute reach-gap effect remained non-significant across specifications, and the main substantive conclusion did not change: participants were only modestly calibrated to the higher-view outcome, while preference and prediction were tightly coupled within the task.

\begin{table*}[t]
\centering
\caption{Primary Logistic Regression Model Predicting Accuracy}
\label{tab:primary_model}
\small
\begin{tabular}{lrrrr}
\toprule
\textbf{Variable} & \textbf{Coefficient} & \textbf{SE} & \textbf{95\% CI} & \textbf{$p$} \\
\midrule
Constant & 0.288 & 0.089 & [0.114, 0.462] & $<.01$ \\
Absolute reach gap (centered) & $-0.006$ & 0.012 & [$-$0.029, 0.017] & .609 \\
Trial index (centered) & 0.014 & 0.009 & [$-$0.004, 0.032] & .123 \\
\midrule
\textit{Category (reference: Animals)} & & & & \\
\quad Comedy & 0.050 & 0.100 & [$-$0.146, 0.247] & .614 \\
\quad Dancing & $-0.512$ & 0.480 & [$-$1.454, 0.429] & .286 \\
\quad Education & 0.186 & 0.266 & [$-$0.337, 0.708] & .486 \\
\quad Meme & $-0.200$ & 0.192 & [$-$0.577, 0.177] & .298 \\
\quad Random & 0.269 & 0.371 & [$-$0.459, 0.997] & .469 \\
\quad Singing & 0.239 & 0.424 & [$-$0.593, 1.070] & .574 \\
\quad Sketch Comedy & $-0.335$ & 0.183 & [$-$0.693, 0.023] & .066 \\
\quad Trend & $-0.304$ & 0.245 & [$-$0.784, 0.176] & .214 \\
\bottomrule
\end{tabular}

\vspace{0.2cm}
\footnotesize
\textit{Note:} Model: $\text{logit}(\Pr(\text{correct} = 1)) = \beta_0 + \beta_1 \cdot \text{abs\_reach\_gap\_c} + \beta_2 \cdot \text{trial\_index\_c} + \sum_{j} \beta_j \cdot \text{category}_j$. Standard errors are two-way clustered by participant (363 clusters) and pair (1978 clusters) using the method of Cameron, Gelbach, and Miller. $N = 3513$ trials from $N = 363$ participants across $N = 1978$ pairs. AIC = 4800.45.
\end{table*}

\begin{table*}
\centering
\caption{Robustness Checks: Alternative Model Specifications}
\label{tab:robustness}
\small
\begin{tabular}{lrrrrr}
\toprule
\textbf{Specification} & \textbf{$N$ trials} & \textbf{$\beta$ (reach gap)} & \textbf{95\% CI} & \textbf{AIC} & \textbf{Notes} \\
\midrule
Base & 3513 & $-0.006$ & [$-$0.029, 0.017] & 4800.45 & Primary specification \\
Signed gap & 3513 & $-0.005$ & [$-$0.028, 0.018] & 4798.35 & Tests directionality effects \\
Exclude trial 1 & 3150 & $-0.003$ & [$-$0.028, 0.021] & 4300.13 & Tests sensitivity to initial anchoring \\
First 2 trials & 715 & $-0.017$ & [$-$0.065, 0.030] & 1001.99 & Early trial robustness (K=2) \\
First 3 trials & 1059 & $-0.009$ & [$-$0.048, 0.030] & 1477.94 & Early trial robustness (K=3) \\
\bottomrule
\end{tabular}

\vspace{0.2cm}
\footnotesize
\textit{Note:} All models use the same formula as the primary model: $\text{correct} \sim \text{abs\_reach\_gap\_c} + \text{trial\_index\_c} + \text{category}$, except ``Signed gap'' which also includes $\text{signed\_reach\_gap\_c}$. Standard errors are two-way clustered by participant and pair. The absolute reach gap coefficient is non-significant across all specifications ($p > .10$ in all cases).
\end{table*}

\subsection{Survey Equivalence and Early-Trial Bounding (RQ4)}

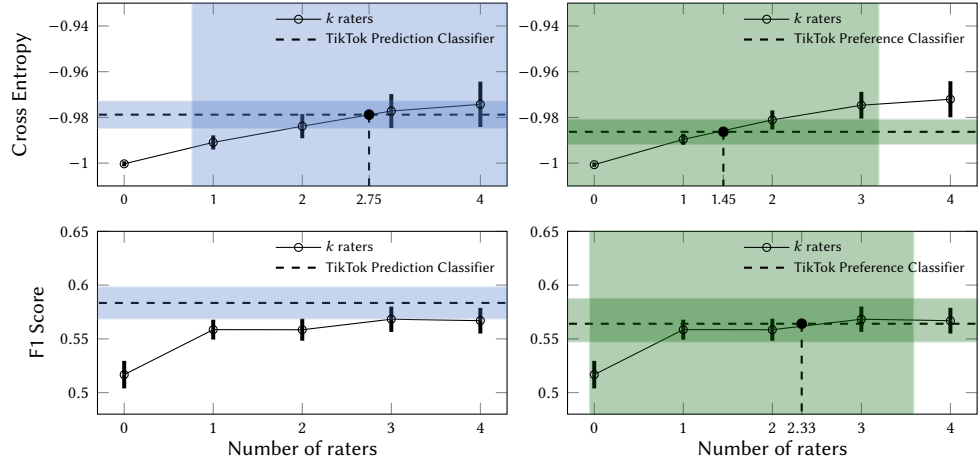
\begin{figure*}[t]
    \centering
    \pgfplotstableread[col sep=comma]{
raters,mean,ci
0,-1.000311,0.001077
1,-0.990949,0.003091
2,-0.983852,0.005275
3,-0.977222,0.007422
4,-0.974287,0.009932
}\preddata

\pgfplotstableread[col sep=comma]{
raters,mean,ci
0,-1.000681,0.000954
1,-0.989625,0.002425
2,-0.981106,0.004163
3,-0.974676,0.005862
4,-0.972057,0.007892
}\prefdata

\pgfplotstableread[col sep=comma]{
raters,mean,ci
0,0.516740,0.012817
1,0.558594,0.009234
2,0.558484,0.010231
3,0.568280,0.011781
4,0.566910,0.011962
}\preddatafone

\pgfplotstableread[col sep=comma]{
raters,mean,ci
0,0.477591,0.014738
1,0.564311,0.006738
2,0.564311,0.007723
3,0.592693,0.011729
4,0.593674,0.012060
}\prefdatafone

\begin{tikzpicture}
\sffamily
\begin{groupplot}[
    group style={
        group size=2 by 2,
        horizontal sep=0.8cm,
        vertical sep=0.6cm,
    }, 
    width=7cm,
    height=4cm,
    xmin=-0.3,xmax=1.3,
    yticklabel style={
    		/pgf/number format/fixed,
    		/pgf/number format/precision=5
    },
    scaled y ticks=false    
]

\nextgroupplot[
legend style={font=\small,
	nodes={scale=0.7, transform shape},
	at={(1,1)},
	anchor=north east,
	draw=none, fill=none},
legend cell align={left},
ylabel = {\small Cross Entropy},
xmin=-0.3,xmax=4.3,ymin=-1.01,ymax=-0.93,
xtick={0, 1, 2, 2.75, 4, 5},
xticklabels={0, 1, 2, 2.75, 4, 5},
xlabel={},
]

\addplot+[
    solid,     
    mark=o,
    draw=black,
    mark options={fill=black, draw=black},
    mark size=1.5pt,
    error bars/.cd,
        y dir=both,
        y explicit,
        error mark=bar,
        error bar style={line width=1.5pt, draw=black}
] table [x=raters, y=mean, y error=ci] {\preddata};

\addlegendentry{$k$ raters}

\addplot[mark=none, black, dashed, thick, samples=2, domain=-1:7] {-0.9788};
\addlegendentry{TikTok Prediction Classifier}

\draw [pptblue!50, fill=pptblue, opacity=0.3] (axis cs:-1,-0.9788-0.006) rectangle (axis cs:31,-0.9788+0.006);

\addplot[mark=*, mark options={scale=.8}, black, thick]
    table[]{
        2.75	-0.9788
    };

\addplot[mark=none, black, dashed, thick, samples=2, domain=-1:31] coordinates {(2.75,-0.9788) (2.75,-2)};

\draw [pptblue!50, fill=pptblue, opacity=0.3] (axis cs:0.76,0) rectangle (axis cs:5.0,-2);

\nextgroupplot[
legend style={font=\small,
	nodes={scale=0.7, transform shape},
	at={(1,1)},
	anchor=north east,
	draw=none, fill=none},
legend cell align={left},
ylabel = {},
xmin=-0.3,xmax=4.3,ymin=-1.01,ymax=-0.93,
xtick={0, 1, 1.45, 2, 3, 4, 5},
xticklabels={0, 1, 1.45, 2, 3, 4, 5},
xlabel={},
]

\addplot+[
    solid,     
    mark=o,
    draw=black,
    mark options={fill=black, draw=black},
    mark size=1.5pt,
    error bars/.cd,
        y dir=both,
        y explicit,
        error mark=bar,
        error bar style={line width=1.5pt, draw=black}
] table [x=raters, y=mean, y error=ci] {\prefdata};

\addlegendentry{$k$ raters}

\addplot[mark=none, black, dashed, thick, samples=2, domain=-1:7] {-0.986311};
\addlegendentry{TikTok Preference Classifier}

\draw [pptgreen!50, fill=pptgreen, opacity=0.3] (axis cs:-1,-0.986311-0.005516) rectangle (axis cs:31,-0.986311+0.005516);

\addplot[mark=*, mark options={scale=.8}, black, thick]
    table[]{
        1.450402	-0.986311
    };

\addplot[mark=none, black, dashed, thick, samples=2, domain=-1:31] coordinates {(1.450402,-0.986311) (1.450402,-2)};

\draw [pptgreen!50, fill=pptgreen, opacity=0.3] (axis cs:-0.297154,0) rectangle (axis cs:3.197957,-2);

\nextgroupplot[
legend style={font=\small,
	nodes={scale=0.7, transform shape},
	at={(1,1)},
	anchor=north east,
	draw=none, fill=none},
legend cell align={left},
ylabel = {\small F1 Score},
xmin=-0.3,xmax=4.3,ymin=0.48,ymax=0.65,
xtick={0, 1, 2, 3, 4, 5},
xlabel={\small Number of raters},
xlabel style = {yshift=0.05in},
]

\addplot+[
    solid,     
    mark=o,
    draw=black,
    mark options={fill=black, draw=black},
    mark size=1.5pt,
    error bars/.cd,
        y dir=both,
        y explicit,
        error mark=bar,
        error bar style={line width=1.5pt, draw=black}
] table [x=raters, y=mean, y error=ci] {\preddatafone};

\addlegendentry{$k$ raters}

\addplot[mark=none, black, dashed, thick, samples=2, domain=-1:7] {0.58342};
\addlegendentry{TikTok Prediction Classifier}

\draw [pptblue!50, fill=pptblue, opacity=0.3] (axis cs:-1, 0.58342-0.015) rectangle (axis cs:7, 0.58342+0.015);




\nextgroupplot[
legend style={font=\small,
	nodes={scale=0.7, transform shape},
	at={(1,1)},
	anchor=north east,
	draw=none, fill=none},
legend cell align={left},
ylabel = {},
xmin=-0.3,xmax=4.3,ymin=0.48,ymax=0.65,
xtick={0, 1, 2, 2.33, 3, 4, 5},
xlabel={\small Number of raters},
xlabel style = {yshift=0.05in},
]

\addplot+[
    solid,     
    mark=o,
    draw=black,
    mark options={fill=black, draw=black},
    mark size=1.5pt,
    error bars/.cd,
        y dir=both,
        y explicit,
        error mark=bar,
        error bar style={line width=1.5pt, draw=black}
] table [x=raters, y=mean, y error=ci] {\preddatafone};

\addlegendentry{$k$ raters}

\addplot[mark=none, black, dashed, thick, samples=2, domain=-1:7] {0.56406};
\addlegendentry{TikTok Preference Classifier}

\draw [pptgreen!50, fill=pptgreen, opacity=0.3] (axis cs:-1, 0.56406-0.017) rectangle (axis cs:7, 0.57073+0.017);

\addplot[mark=*, mark options={scale=.8}, black, thick]
    table[]{
        2.33    0.56406
    };

\addplot[mark=none, black, dashed, thick, samples=2, domain=-1:31] coordinates {(2.33,0.56406) (2.33,0)};

\draw [pptgreen!50, fill=pptgreen, opacity=0.3] (axis cs:-0.054,1) rectangle (axis cs:3.59,-1);

\end{groupplot}
\end{tikzpicture}
    \caption{Survey equivalence comparison of TikTok's view-based signal against human judgments. Each panel shows how performance scales with the number of human raters (1 to 5), with TikTok's signal plotted as a constant reference line. Top panels report cross-entropy; bottom panels report F1 score. Left panels correspond to participants' popularity predictions, and right panels to their personal preferences.}
    \label{fig:surveyequiv}
\end{figure*}

To answer \textbf{RQ4}, we use Survey Equivalence to translate alignment into a more interpretable unit: how many independent human raters provide a signal comparable to TikTok's view-based outcome. In this framework, the view-based signal is treated as an additional classifier and compared against aggregated human judgments as the number of human raters increases. Under this evaluation, TikTok's view-based outcome aligned with aggregated human judgments at a level comparable to roughly two human raters (Fig.~\ref{fig:surveyequiv}).

This result reinforces the broader descriptive picture of the paper. The public-view benchmark is not wildly disconnected from human judgment, but neither does it line up with a large or especially reliable crowd signal in this cue-limited setting. The correspondence is limited enough to support a narrower interpretation of outcome legibility: participants were not simply reading realized reach directly from content.

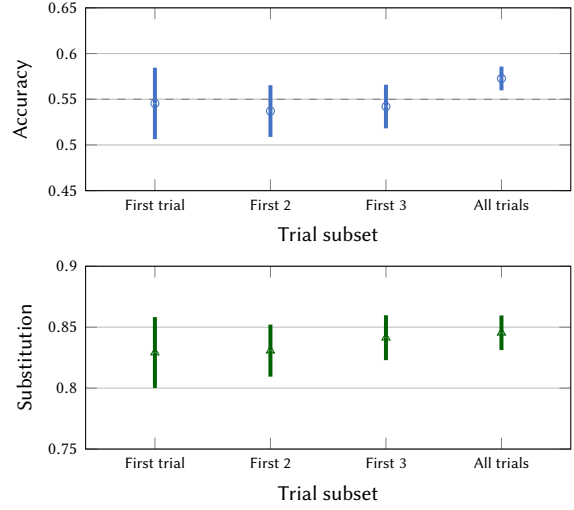
\begin{figure}[t]
    \centering
\pgfplotstableread[col sep=comma]{
subset,Accuracy,AccuracyCI,Substitution,SubstitutionCI
First trial only,0.545455,0.039056,0.829201,0.029072
First 2 trials,0.537063,0.028272,0.830769,0.021346
First 3 trials,0.542021,0.023856,0.84136,0.018415
All trials,0.57273,0.013004,0.845431,0.014163
}\earlytrialdata

\begin{tikzpicture}
\begin{groupplot}[
    group style={
        group size=1 by 2,
        horizontal sep=0.2cm,
        vertical sep=1cm,
    }, 
    ymajorgrids=true,
]

\nextgroupplot[
    width=8cm,
    height=4cm,
    ylabel={Accuracy},
    ymin=0.45, ymax=0.65,
    enlarge x limits=0.2,
    xtick={0,1,2,3},
    xticklabels={First trial,First 2,First 3,All trials},
    xlabel={Trial subset},
    legend style={at={(1.0,1.0)}, anchor=north east, legend columns=1, draw=none, fill=none, font=\sffamily\scriptsize},
]

\addplot+[
    only marks,
    mark=o,
    mark options={fill=pptblue, draw=pptblue},
    mark size=1.5pt,
    error bars/.cd,
        y dir=both,
        y explicit,
        error mark=bar,
        error bar style={line width=1.5pt, draw=pptblue},
] table [
    x expr=\coordindex, 
    y=Accuracy, 
    y error=AccuracyCI,
] {\earlytrialdata};

\draw[dashed, gray] ({rel axis cs:0,0.5}) -- ({rel axis cs:1,0.5});

\nextgroupplot[
    width=8cm,
    height=4cm,
    ylabel={Substitution},
    ymin=0.75, ymax=0.90,
    enlarge x limits=0.2,
    xtick={0,1,2,3},
    xticklabels={First trial,First 2,First 3,All trials},
    xlabel={Trial subset},
]

\addplot+[
    only marks,
    mark=triangle,
    mark options={fill=pptgreen, draw=pptgreen},
    mark size=1.75pt,
    error bars/.cd,
        y dir=both,
        y explicit,
        error mark=bar,
        error bar style={line width=1.5pt, draw=pptgreen},
] table [
    x expr=\coordindex, 
    y=Substitution, 
    y error=SubstitutionCI,
] {\earlytrialdata};

\end{groupplot}
\end{tikzpicture}
    \caption{Early-trial bounding analysis. \textbf{Top:} Accuracy rates for first trial only, first 2 trials, first 3 trials, and all trials. \textbf{Bottom:} Preference--prediction matching rates for the same trial subsets. Error bars indicate 95\% participant-cluster bootstrap confidence intervals. The dashed line in the top panel marks chance level (50\%).}
    \label{fig:early_trials}
\end{figure}

We also used early-trial bounding analyses to test whether the main patterns were driven by across-trial learning or feedback drift. Accuracy and preference--prediction matching were similar in the first trial(s) and in the full set of trials (Fig.~\ref{fig:early_trials}), which suggests that the core patterns were present from the outset rather than emerging gradually over repeated play. This is useful as a bound on across-trial drift, but it does not resolve the stronger validity concern raised by the fixed within-trial question order. The early-trial analyses therefore narrow one alternative explanation without ruling out within-trial anchoring.

\subsection{Participant-Level Patterns (Exploratory)}

To provide descriptive context on individual differences, we asked participants seven survey questions about TikTok usage habits, tenure, and engagement behaviors. We then fit a participant-level model using average survey responses to predict correctness. Most predictors were small and non-significant, so we treat this analysis as exploratory rather than central to the paper's main claims.

\begin{figure}[t]
    \centering
    \pgfplotstableread[col sep=comma]{
x,term,oddsratio,ci95
1,q1_ordinal,1.0225,0.0235
2,q2_ordinal,1.0175,0.0236
3,q3_binary,0.9854,0.0343
4,q4_ordinal,1.0134,0.0302
5,q5_ordinal,0.9868,0.0274
6,q6_ordinal,0.9857,0.0264
7,q7_ordinal,0.9749,0.0250
}\logoddsdata

\pgfplotstableread[col sep=comma]{
x,term,oddsratio,ci95
1,q1_ordinal,1.0168,0.0242
2,q2_ordinal,1.0215,0.0245
3,q3_binary,0.9830,0.0354
4,q4_ordinal,0.9872,0.0304
5,q5_ordinal,0.9970,0.0286
6,q6_ordinal,1.0007,0.0277
7,q7_ordinal,0.9740,0.0258
}\logoddsdatapref

\begin{tikzpicture}
\begin{axis}[
    xlabel={Odds Ratio (log scale)},
    ylabel={Question},
    y dir=reverse,
    ymin=0, ymax=6,
    xmin=0.9, xmax=1.1,
    ytick={0,1,2,3,4,5,6},
    yticklabels={Use Per Month, How long, Like vs View Count, Consider Likes, Consider Views, Give Likes, Make Comments},
    width=7cm,
    height=5cm,
    enlarge y limits=0.1,
    legend style={at={(0.0,1.0)}, anchor=north west, legend columns=1, draw=none, fill=none, font=\sffamily\scriptsize},
    legend cell align={left},    
]

\addplot+[
    only marks,
    mark=o,
    mark options={fill=pptblue, draw=pptblue},
    mark size=1.75pt,
    error bars/.cd,
        x dir=both,
        x explicit,
        error mark=bar,
        error bar style={line width=1.5pt, draw=pptblue},
] table[
    x=oddsratio,
    y expr=\coordindex + 0.1,
    x error=ci95
] \logoddsdata;

\addlegendentry{Pred. Accuracy}

\addplot+[
    only marks,
    mark=triangle,
    mark options={fill=pptgreen, draw=pptgreen},
    mark size=2.5pt,
    error bars/.cd,
        x dir=both,
        x explicit,
        error mark=bar,
        error bar style={line width=1.5pt, draw=pptgreen},
] table[
    x=oddsratio,
    y expr=\coordindex - 0.1,
    x error=ci95
] \logoddsdatapref;

\addlegendentry{Pref. Alignment}

\addplot[dashed, gray] coordinates {(1.0,-0.15) (1.0,6.5)};

\end{axis}
\end{tikzpicture}
    \caption{Participant-level model (odds ratios with 95\% CIs) for predicting correct identification. Odds ratios greater than 1 indicate positive relationships.}
    \label{fig:logodds}
\end{figure}
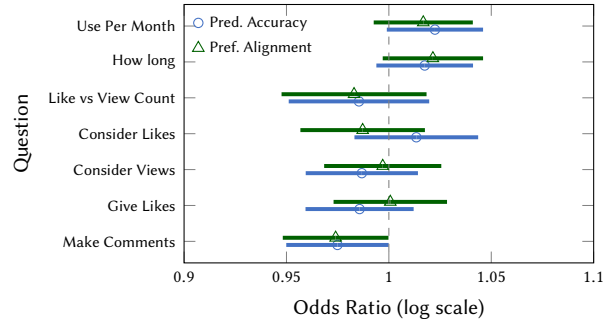

Two descriptive patterns were the most notable. More frequent TikTok use was associated with somewhat better performance, while more frequent commenting was associated with lower accuracy. Daily TikTok users showed higher accuracy (60.5\% correct) and greater preference--reach alignment (58.8\%) than non-users (57.3\% correct, 55.9\% alignment). Participants with three or more years of experience achieved 60.2\% accuracy compared to 52.0\% among those with three to twelve months of use. By contrast, self-reported attention to metrics showed only small differences (e.g., 58.5\% vs.\ 56.1\% for like-focused vs.\ view-focused participants). These patterns suggest that tacit familiarity with the platform may matter more than explicit self-reported attention to metrics, but the evidence here is descriptive and should be interpreted cautiously (Fig.~\ref{fig:logodds}).

\section{Discussion}

\subsection{Interpreting Reach under Reduced Metrics}

When platforms hide or downplay visible counters, users do not stop interpreting what they see. They still form expectations about what is spreading, what the platform seems to reward, and what kinds of content count as successful. Our results suggest that these judgments are only loosely calibrated to realized platform reach when explicit popularity cues are unavailable. Participants identified the higher-view video only modestly above chance, while their preferences and reach judgments aligned much more often within the task. Taken together, these findings do not suggest that users can reliably read reach from content alone. They instead point to a more uncertain interpretive process in which judgments about reach often travel through personal taste and other weak cues.

That narrower conclusion matters for how we think about \emph{outcome legibility}. In this paper, the concept is most useful not as a broad account of how people understand algorithmic systems in general, but as a way of naming a more specific problem: how users interpret visible outcomes of platform circulation when the most obvious counters are muted. Our evidence speaks directly to that bounded question. It shows that people still make reach judgments under hidden metrics, but those judgments do not line up strongly with the observed view-based outcome. For this reason, we treat the task less as a test of whether users can reverse-engineer TikTok and more as a window into what kinds of signals remain available when platforms reduce visible metrics.

The results also help sharpen the relationship between this paper and earlier ``guess the crowd'' work. As in GuessTheKarma, the central empirical picture is not one of strong collective legibility, but of only partial correspondence between engagement-derived outcomes and what people judge from content in a metric-hidden setting. What TikTok adds is a recommendation-driven short-video environment where genre convention, trend recognition, and cultural fit may matter alongside more familiar production cues. The contribution, then, is not that users can cleanly identify successful content on TikTok, but that the platform setting reveals how quickly reach judgments can become entangled with personal and genre-shaped expectations once explicit counters recede.

\subsection{Bounded Orientation as a Design Implication}

One implication of these findings is that removing counters does not remove interpretation. It changes the evidence users have available when they try to understand success. That is where we see value in what we call \emph{bounded orientation}: lightweight, contextual support that helps users interpret reach outcomes without simply restoring precise metrics. The goal is not to tell users exactly why a video performed as it did, and not to recreate a fully transparent system that platforms do not actually provide. It is to reduce the interpretive vacuum that can emerge when counters disappear but judgments about popularity remain socially important.

In practical terms, this could include coarse or delayed ranges rather than precise counts, contextual reminders that reach is contingent on timing and audience overlap, or genre-aware prompts that make outcomes feel less like stable properties of the content itself. These are modest interventions. They would not change the underlying mechanics of amplification, nor would they solve the broader social influence problems associated with visible metrics. They are better understood as interpretive supports: ways of helping users orient to outcomes that are inherently noisy, contingent, and only partially legible from the interface.

We see this as a middle position between two unsatisfying extremes. On one side, precise counters can invite comparison, herding, and cascade effects. On the other, removing those counters entirely may push users toward idiosyncratic heuristics, overconfidence in personal taste, or misplaced assumptions about suppression and favoritism. Our results do not show that bounded orientation would fix those problems, and we do not claim that this study directly evaluates such interventions. They do, however, suggest why some intermediate layer of interpretive support may be worth considering.

\subsection{Limitations and Threats to Validity}

Several limitations should shape how these findings are interpreted.

\noindent\textbf{Fixed question order and within-trial anchoring.}
The most important limitation is the task structure itself. Within each trial, participants always reported preference before predicting reach. That design means the high rate of preference--prediction matching cannot be taken as clean evidence that users naturally rely on preference when reasoning about reach. It is equally plausible that the ordering introduced anchoring or consistency pressure within the task. Our early-trial analyses and trial-index terms help bound \emph{across-trial} learning or feedback drift, but they do not resolve the stronger validity concern of \emph{within-trial} anchoring. A stronger design would randomize or separate these judgments.

\noindent\textbf{Reach proxy scope.}
We benchmark judgments against verified public view counts, using them as a bounded proxy for realized platform reach. This is useful because it provides a concrete and observable outcome, but it is not a pure measure of algorithmic exposure. View totals may also reflect creator audience, video age, prior notoriety, and off-platform traffic. The results should therefore be read as calibration to a public attention outcome, not as a direct test of whether users can perceive TikTok's internal ranking logic.

\noindent\textbf{Task realism and ecological scope.}
\textit{TokOrNot} is a study instrument, not a simulation of the For You Page. TikTok is ordinarily experienced as a sequential, personalized feed rather than a side-by-side comparison task. The paired design is useful because it holds exposure constant and makes comparative judgments analyzable, but it also strips away important features of platform use such as scrolling context, repeated account exposure, comment cues, and ambient familiarity with creators or trends. The task therefore captures one bounded form of reach inference under hidden metrics rather than everyday platform interpretation in full.

\noindent\textbf{Sample and generalizability.}
Our participant pool consisted of U.S.-based Prolific adults with a mean age substantially older than TikTok's core user base. That matters because platform familiarity, generational media habits, and cultural proximity to TikTok genres may all shape what people treat as cues of likely reach. For this reason, we avoid claiming that the observed rates directly characterize TikTok users as a whole. A younger or more globally distributed sample might reason differently about the same pairs.

\noindent\textbf{Limited direct leverage on metric-hiding policy.}
Finally, the paper speaks to a design problem created by reduced metrics, but it does not directly evaluate a metric-hiding intervention. We do not compare visible-count and hidden-count conditions, nor do we test concrete interface alternatives. The study is better understood as diagnosing how people reason in a cue-limited environment than as establishing the consequences of any specific platform policy.

\section{Conclusion}

When visible metrics recede, users do not stop making judgments about what is popular. They continue to infer which videos reached broader audiences, drawing on personal preference, qualitative cues, and informal theories about platform and audience dynamics. In our pairwise TikTok task, those judgments were only modestly aligned with the higher-view outcome, while preference and prediction aligned much more strongly within the task. Taken together, these findings suggest that realized platform reach is not easily legible from content alone when explicit popularity signals are unavailable.

That is the main contribution of this paper. Rather than treating users' judgments as a window into TikTok's internal ranking logic, we benchmark them against an observable outcome of platform circulation: realized platform reach proxied by public view counts. This outcome-centered framing keeps the analysis focused on what users can actually see and infer from the interface, even when the mechanisms shaping distribution remain opaque. The Survey Equivalence analysis extends that contribution by offering a more interpretable bridge between platform outcomes and small-group human judgments.

The broader implication is that metric reduction does not eliminate interpretation. It changes what interpretation is built from. When platforms hide or soften visible counters, users still try to understand what success looks like, but they may do so using weaker, more personal, and more uneven cues. That remains an important challenge for social computing research and practice, especially for understanding how platforms shape users' expectations of what ``popular'' means on-platform.

Several directions follow from this work. Future studies could test how these patterns change across other content regimes, including influencer content, branded videos, or regionally specific trends, where folk theories of reach may operate differently. More diverse samples, especially younger and more internationally distributed users, would help clarify how platform familiarity and cultural context shape these judgments. Methodologically, follow-up studies could randomize question order, compare hidden- and visible-metric conditions directly, or use behavioral measures to better understand how participants arrive at their decisions. Design-oriented work could also evaluate whether coarse metric ranges, delayed visibility, or other lightweight interpretive supports help users reason about reach without simply reintroducing the social influence effects of precise counters.

\bibliographystyle{ACM-Reference-Format}
\bibliography{TokOrNot}

@misc{BandyDiakopoulos-TulsaFlopCaseStudy-2020,
  title = {\#{{TulsaFlop}}: {{A Case Study}} of {{Algorithmically-Influenced Collective Action}} on {{TikTok}}},
  shorttitle = {\#{{TulsaFlop}}},
  author = {Bandy, Jack and Diakopoulos, Nicholas},
  year = {2020},
  month = dec,
  number = {arXiv:2012.07716},
  eprint = {2012.07716},
  primaryclass = {cs},
  publisher = {arXiv},
  doi = {10.48550/arXiv.2012.07716},
  url = {http://arxiv.org/abs/2012.07716},
  urldate = {2023-09-24},
  archiveprefix = {arXiv}
}

@incollection{CollieWilson-Barnao-PlayingTikTokAlgorithmic-2020,
  title = {Playing with {{TikTok}}: Algorithmic Culture and the Future of Creative Work},
  shorttitle = {Playing with {{TikTok}}},
  booktitle = {The {{Future}} of {{Creative Work}}},
  author = {Collie, Natalie and {Wilson-Barnao}, Caroline},
  year = {2020},
  month = sep,
  pages = {172--188},
  publisher = {Edward Elgar Publishing},
  url = {https://china.elgaronline.com/edcollchap/edcoll/9781839101090/9781839101090.00020.xml},
  urldate = {2023-10-04},
  chapter = {The Future of Creative Work},
  isbn = {978-1-83910-110-6},
  langid = {american}
}

@article{Darvin-DesignResistancePerformance-2022,
  title = {Design, Resistance and the Performance of Identity on {{TikTok}}},
  author = {Darvin, Ron},
  year = {2022},
  month = apr,
  journal = {Discourse, Context \& Media},
  volume = {46},
  pages = {100591},
  issn = {2211-6958},
  doi = {10.1016/j.dcm.2022.100591},
  url = {https://www.sciencedirect.com/science/article/pii/S2211695822000149},
  urldate = {2023-09-19}
}

@inproceedings{DeVitoEtAl-AlgorithmsRuinEverything-2017,
  title = {"{{Algorithms}} Ruin Everything": \#{{RIPTwitter}}, {{Folk Theories}}, and {{Resistance}} to {{Algorithmic Change}} in {{Social Media}}},
  shorttitle = {"{{Algorithms}} Ruin Everything"},
  author = {DeVito, Michael A. and Gergle, Darren and Birnholtz, Jeremy},
  year = {2017},
  month = may,
  series = {{{CHI}} '17},
  pages = {3163--3174},
  publisher = {Association for Computing Machinery},
  address = {New York, NY, USA},
  doi = {10.1145/3025453.3025659},
  url = {https://dl.acm.org/doi/10.1145/3025453.3025659},
  urldate = {2024-02-27},
  isbn = {978-1-4503-4655-9}
}

@inproceedings{DeVitoEtAl-HowPeopleForm-2018,
  title = {How {{People Form Folk Theories}} of {{Social Media Feeds}} and {{What}} It {{Means}} for {{How We Study Self-Presentation}}},
  author = {DeVito, Michael A. and Birnholtz, Jeremy and Hancock, Jeffery T. and French, Megan and Liu, Sunny},
  year = {2018},
  month = apr,
  series = {{{CHI}} '18},
  pages = {1--12},
  publisher = {Association for Computing Machinery},
  address = {New York, NY, USA},
  doi = {10.1145/3173574.3173694},
  url = {https://dl.acm.org/doi/10.1145/3173574.3173694},
  urldate = {2024-02-27},
  isbn = {978-1-4503-5620-6}
}

@inproceedings{EslamiEtAl-AlwaysAssumedThat-2015,
  title = {"{{I}} Always Assumed That {{I}} Wasn't Really That Close to [Her]": {{Reasoning}} about {{Invisible Algorithms}} in {{News Feeds}}},
  shorttitle = {"{{I}} Always Assumed That {{I}} Wasn't Really That Close to [Her]"},
  booktitle = {Proceedings of the 33rd {{Annual ACM Conference}} on {{Human Factors}} in {{Computing Systems}}},
  author = {Eslami, Motahhare and Rickman, Aimee and Vaccaro, Kristen and Aleyasen, Amirhossein and Vuong, Andy and Karahalios, Karrie and Hamilton, Kevin and Sandvig, Christian},
  year = {2015},
  month = apr,
  series = {{{CHI}} '15},
  pages = {153--162},
  publisher = {Association for Computing Machinery},
  address = {New York, NY, USA},
  doi = {10.1145/2702123.2702556},
  url = {https://dl.acm.org/doi/10.1145/2702123.2702556},
  urldate = {2025-05-06},
  isbn = {978-1-4503-3145-6}
}

@inproceedings{JacoviEtAl-PerceptionOthersInferring-2014,
  title = {The Perception of Others: Inferring Reputation from Social Media in the Enterprise},
  shorttitle = {The Perception of Others},
  booktitle = {Proceedings of the 17th {{ACM}} Conference on {{Computer}} Supported Cooperative Work \& Social Computing},
  author = {Jacovi, Michal and Guy, Ido and {Kremer-Davidson}, Shiri and Porat, Sara and {Aizenbud-Reshef}, Netta},
  year = {2014},
  month = feb,
  series = {{{CSCW}} '14},
  pages = {756--766},
  publisher = {Association for Computing Machinery},
  address = {New York, NY, USA},
  doi = {10.1145/2531602.2531667},
  url = {https://doi.org/10.1145/2531602.2531667},
  urldate = {2025-05-13},
  isbn = {978-1-4503-2540-0}
}

@article{LorenzEtAl-HowSocialInfluence-2011,
  title = {How Social Influence Can Undermine the Wisdom of Crowd Effect},
  author = {{Jan Lorenz} and Lorenz, Jan and {Heiko Rauhut} and Rauhut, Heiko and {Frank Schweitzer} and Schweitzer, Frank and {Dirk Helbing} and Helbing, Dirk},
  year = {2011},
  month = may,
  journal = {Proceedings of the National Academy of Sciences of the United States of America},
  volume = {108},
  number = {22},
  pages = {9020--9025},
  doi = {10.1073/pnas.1008636108},
  pmcid = {3107299},
  pmid = {21576485}
}

@article{KarizatEtAl-AlgorithmicFolkTheories-2021,
  title = {Algorithmic {{Folk Theories}} and {{Identity}}: {{How TikTok Users Co-Produce Knowledge}} of {{Identity}} and {{Engage}} in {{Algorithmic Resistance}}},
  shorttitle = {Algorithmic {{Folk Theories}} and {{Identity}}},
  author = {Karizat, Nadia and Delmonaco, Dan and Eslami, Motahhare and Andalibi, Nazanin},
  year = {2021},
  month = oct,
  journal = {Proceedings of the ACM on Human-Computer Interaction},
  volume = {5},
  number = {CSCW2},
  pages = {305:1--305:44},
  doi = {10.1145/3476046},
  url = {https://dl.acm.org/doi/10.1145/3476046},
  urldate = {2024-02-27}
}

@article{LevMuchnikEtAl-SocialInfluenceBias-2013,
  title = {Social {{Influence Bias}}: {{A Randomized Experiment}}},
  author = {{Lev Muchnik} and Muchnik, Lev and {Sinan Aral} and Aral, Sinan and {Sean J. Taylor} and Taylor, Sean J.},
  year = {2013},
  month = aug,
  journal = {Science},
  volume = {341},
  number = {6146},
  pages = {647--651},
  doi = {10.1126/science.1240466},
  pmid = {23929980}
}

@inproceedings{MarcuEtAl-DesigningCollaborativeReflection-2014,
  title = {Designing for {{Collaborative Reflection}}},
  booktitle = {Proceedings of the 8th {{International Conference}} on {{Pervasive Computing Technologies}} for {{Healthcare}}},
  author = {Marcu, Gabriela and Dey, Anind and Kiesler, Sara},
  year = {2014},
  publisher = {ICST},
  address = {Oldenburg, Germany},
  doi = {10.4108/icst.pervasivehealth.2014.254987},
  url = {http://eudl.eu/doi/10.4108/icst.pervasivehealth.2014.254987},
  urldate = {2025-05-07},
  isbn = {978-1-63190-011-2},
  langid = {english}
}

@article{MariaGlenskiEtAl-GuessTheKarmaGameAssess-2018,
  title = {{{GuessTheKarma}}: {{A Game}} to {{Assess Social Rating Systems}}},
  author = {Glenski, Maria and Stoddard, Greg and Resnick, Paul and Weninger, Tim},
  year = {2018},
  month = sep,
  journal = {arXiv: Human-Computer Interaction},
  doi = {10.1145/3274328}
}

@misc{Matsa-MoreAmericansAre-2023,
  title = {More {{Americans}} Are Getting News on {{TikTok}}, Bucking the Trend on Other Social Media Sites},
  author = {Matsa, Katerina Eva},
  year = {2023},
  month = nov,
  journal = {Pew Research Center},
  url = {https://www.pewresearch.org/short-reads/2022/10/21/more-americans-are-getting-news-on-tiktok-bucking-the-trend-on-other-social-media-sites/},
  urldate = {2023-09-25},
  langid = {american}
}

@article{SalganikEtAl-ExperimentalStudyInequality-2006,
  title = {Experimental {{Study}} of {{Inequality}} and {{Unpredictability}} in an {{Artificial Cultural Market}}},
  author = {{Matthew J. Salganik} and Salganik, Matthew J. and {Peter Sheridan Dodds} and Dodds, Peter Sheridan and {Duncan J. Watts} and Watts, Duncan J.},
  year = {2006},
  month = feb,
  journal = {Science},
  volume = {311},
  number = {5762},
  pages = {854--856},
  doi = {10.1126/science.1121066},
  pmid = {16469928}
}

@inproceedings{MuralikumarBietz-VisualizingAlgorithmicSelection-2019,
  title = {Visualizing {{Algorithmic Selection}} in {{Social Media}}},
  booktitle = {Companion {{Publication}} of the 2019 {{Conference}} on {{Computer Supported Cooperative Work}} and {{Social Computing}}},
  author = {Muralikumar, Meena Devii and Bietz, Matthew J.},
  year = {2019},
  month = nov,
  series = {{{CSCW}} '19 {{Companion}}},
  pages = {319--323},
  publisher = {Association for Computing Machinery},
  address = {New York, NY, USA},
  doi = {10.1145/3311957.3359476},
  url = {https://doi.org/10.1145/3311957.3359476},
  urldate = {2025-05-13},
  isbn = {978-1-4503-6692-2}
}

@inproceedings{VaccaroEtAl-ContestabilityAlgorithmicSystems-2019,
  title = {Contestability in {{Algorithmic Systems}}},
  booktitle = {Companion {{Publication}} of the 2019 {{Conference}} on {{Computer Supported Cooperative Work}} and {{Social Computing}}},
  author = {Vaccaro, Kristen and Karahalios, Karrie and Mulligan, Deirdre K. and Kluttz, Daniel and Hirsch, Tad},
  year = {2019},
  month = nov,
  series = {{{CSCW}} '19 {{Companion}}},
  pages = {523--527},
  publisher = {Association for Computing Machinery},
  address = {New York, NY, USA},
  doi = {10.1145/3311957.3359435},
  url = {https://dl.acm.org/doi/10.1145/3311957.3359435},
  urldate = {2025-05-13},
  isbn = {978-1-4503-6692-2}
}

@article{VelkovaKaun-AlgorithmicResistanceMedia-2021,
  title = {Algorithmic Resistance: Media Practices and the Politics of Repair},
  shorttitle = {Algorithmic Resistance},
  author = {Velkova, Julia and Kaun, Anne},
  year = {2021},
  month = mar,
  journal = {Information, Communication \& Society},
  volume = {24},
  number = {4},
  pages = {523--540},
  issn = {1369-118X},
  doi = {10.1080/1369118X.2019.1657162},
  url = {https://doi.org/10.1080/1369118X.2019.1657162},
  urldate = {2024-02-27}
}

@inproceedings{VombatkereEtAl-TikTokArtPersonalization-2024,
  title = {{{TikTok}} and the {{Art}} of {{Personalization}}: {{Investigating Exploration}} and {{Exploitation}} on {{Social Media Feeds}}},
  shorttitle = {{{TikTok}} and the {{Art}} of {{Personalization}}},
  booktitle = {Proceedings of the {{ACM Web Conference}} 2024},
  author = {Vombatkere, Karan and Mousavi, Sepehr and Zannettou, Savvas and Roesner, Franziska and Gummadi, Krishna P.},
  year = {2024},
  month = may,
  series = {{{WWW}} '24},
  pages = {3789--3797},
  publisher = {Association for Computing Machinery},
  address = {New York, NY, USA},
  doi = {10.1145/3589334.3645600},
  url = {https://dl.acm.org/doi/10.1145/3589334.3645600},
  urldate = {2025-05-06},
  isbn = {9798400701719}
}

@inproceedings{WangEtAl-QuantifyingHerdingEffects-2014,
  title = {Quantifying Herding Effects in Crowd Wisdom},
  booktitle = {Proceedings of the 20th {{ACM SIGKDD}} International Conference on {{Knowledge}} Discovery and Data Mining},
  author = {Wang, Ting and Wang, Dashun and Wang, Fei},
  year = {2014},
  month = aug,
  series = {{{KDD}} '14},
  pages = {1087--1096},
  publisher = {Association for Computing Machinery},
  address = {New York, NY, USA},
  doi = {10.1145/2623330.2623720},
  url = {https://dl.acm.org/doi/10.1145/2623330.2623720},
  urldate = {2025-05-06},
  isbn = {978-1-4503-2956-9}
}

@inproceedings{WuHuberman-HowPublicOpinion-2008,
  title = {How {{Public Opinion Forms}}},
  booktitle = {Internet and {{Network Economics}}},
  author = {Wu, Fang and Huberman, Bernardo A.},
  editor = {Papadimitriou, Christos and Zhang, Shuzhong},
  year = {2008},
  pages = {334--341},
  publisher = {Springer},
  address = {Berlin, Heidelberg},
  doi = {10.1007/978-3-540-92185-1_39},
  isbn = {978-3-540-92185-1},
  langid = {english}
}

@article{ZulliZulli-ExtendingInternetMeme-2020,
  title = {Extending the {{Internet}} Meme: {{Conceptualizing}} Technological Mimesis and Imitation Publics on the {{TikTok}} Platform},
  shorttitle = {Extending the {{Internet}} Meme},
  author = {Zulli, Diana and Zulli, David James},
  year = {2020},
  month = dec,
  journal = {New Media \& Society},
  volume = {24},
  number = {8},
  pages = {1872--1890},
  publisher = {SAGE Publications},
  issn = {1461-4448},
  doi = {10.1177/1461444820983603},
  url = {https://doi.org/10.1177/1461444820983603},
  urldate = {2024-03-02},
  langid = {english}
}

@article{wit1998rational,
  title={Rational choice and the Condorcet jury theorem},
  author={Wit, J{\"o}rgen},
  journal={Games and Economic Behavior},
  volume={22},
  number={2},
  pages={364--376},
  year={1998},
  publisher={Elsevier}
}

@article{MaywormEtAl-ContentModerationFolkTheories-2024,
  author = {Samuel Mayworm and Michael Ann DeVito and Daniel Delmonaco and Hibby Thach and Oliver L. Haimson},
  title = {Content Moderation Folk Theories and Perceptions of Platform Spirit among Marginalized Social Media Users},
  year = {2024},
  journal = {ACM Transactions on Social Computing},
  volume = {7},
  number = {1--4},
  articleno = {1},
  numpages = {27},
  doi = {10.1145/3632741},
  url = {https://doi.org/10.1145/3632741},
  publisher = {Association for Computing Machinery},
  address = {New York, NY, USA}
}

@article{DelmonacoEtAl-WhatAreYouDoingTikTok-2024,
  author = {Daniel Delmonaco and Samuel Mayworm and Hibby Thach and Josh Guberman and Aurelia Augusta and Oliver L. Haimson},
  title = {{\textquotedblleft}What are you doing, TikTok?{\textquotedblright}: How Marginalized Social Media Users Perceive, Theorize, and {\textquotedblleft}Prove{\textquotedblright} Shadowbanning},
  year = {2024},
  journal = {Proceedings of the ACM on Human-Computer Interaction},
  volume = {8},
  number = {CSCW1},
  articleno = {154},
  numpages = {39},
  doi = {10.1145/3637431},
  url = {https://doi.org/10.1145/3637431},
  publisher = {Association for Computing Machinery},
  address = {New York, NY, USA}
}

@article{BartaAndalibi-TheorizingSelfVisibility-2024,
  author = {Kristen Barta and Nazanin Andalibi},
  title = {Theorizing Self Visibility on Social Media: A Visibility Objects Lens},
  year = {2024},
  journal = {ACM Transactions on Computer-Human Interaction},
  volume = {31},
  number = {3},
  articleno = {31},
  numpages = {28},
  doi = {10.1145/3660337},
  url = {https://doi.org/10.1145/3660337},
  publisher = {Association for Computing Machinery},
  address = {New York, NY, USA}
}

@inproceedings{VeraGhosh-ControllingUnwantedContentTikTok-2025,
  author = {Julie A. Vera and Sourojit Ghosh},
  title = {{\textquotedblleft}They've Over-Emphasized That One Search{\textquotedblright}: Controlling Unwanted Content on TikTok's For You Page},
  year = {2025},
  booktitle = {Proceedings of the 2025 CHI Conference on Human Factors in Computing Systems},
  articleno = {221},
  numpages = {8},
  doi = {10.1145/3706598.3713666},
  url = {https://doi.org/10.1145/3706598.3713666},
  publisher = {Association for Computing Machinery},
  address = {New York, NY, USA}
}

@inproceedings{ChoiEtAl-CreatorFriendlyAlgorithms-2023,
  author = {Yoonseo Choi and Eun Jeong Kang and Min Kyung Lee and Juho Kim},
  title = {Creator-friendly Algorithms: Behaviors, Challenges, and Design Opportunities in Algorithmic Platforms},
  year = {2023},
  booktitle = {Proceedings of the 2023 CHI Conference on Human Factors in Computing Systems},
  articleno = {564},
  numpages = {22},
  doi = {10.1145/3544548.3581386},
  url = {https://doi.org/10.1145/3544548.3581386},
  publisher = {Association for Computing Machinery},
  address = {New York, NY, USA}
}

@inproceedings{KlugEtAl-AlgorithmAwarenessAlgospeak-2023,
  author = {Daniel Klug and Ella Steen and Kathryn Yurechko},
  title = {How Algorithm Awareness Impacts Algospeak Use on TikTok},
  year = {2023},
  booktitle = {Companion Proceedings of the ACM Web Conference 2023},
  pages = {234--237},
  doi = {10.1145/3543873.3587355},
  url = {https://doi.org/10.1145/3543873.3587355},
  publisher = {Association for Computing Machinery},
  address = {New York, NY, USA}
}

\end{document}